\documentstyle{article}
\begin{document}
%
\title{Competition between crystalline electric field singlet 
and itinerant states of f electrons}
\author{Shinji Watanabe\thanks{e-mail:
watanabe@cmpt01.phys.tohoku.ac.jp}\  and Yoshio Kuramoto \\
Department of Physics, Tohoku University, Sendai 980}

\date{\today}
\maketitle

\begin{abstract}
{
A new kind of phase transition is proposed for lattice fermion systems
with simplified $f^{2}$ configurations at each site. 
The free energy of the model is computed in the mean-field
approximation for both the itinerant state with the Kondo screening,
and a localized state with the crystalline electric field (CEF)
singlet at each site.
The presence of a first-order phase transition is demonstrated in
which the itinerant state changes into the localized state toward
lower temperatures.  In the half-filled case,  the insulating
state at high temperatures changes into a metallic state, in marked
contrast with the Mott transition in the Hubbard model.  
For comparison, corresponding states are discussed for the
two-impurity Kondo system with $f^{1}$ configuration at each site. 
}
\end{abstract}

\section{Introduction}\label{chap:1}

In some uranium compounds with $5f^{2}$ configuration 
(${\rm U}^{4+}$) the CEF ground-state can be a nonmagnetic singlet. 
The CEF singlet is also realized in some praseodymium compounds 
with $4f^{2}$ configuration (${\rm Pr}^{3+}$). 
In these cases the spin entropy of the system can go to zero as
temperature 
 decreases even though interactions with conduction electrons or 
with f electrons at other sites are absent.  
This is in striking contrast with the case of cerium compounds with 
$4f^{1}$ configuration (${\rm Ce}^{3+}$); 
the entropy does not disappear 
at zero temperature if a Ce ion is isolated 
because of the Kramers degeneracy associated with the $f^{1}$
configuration. 
As a result the system chooses, depending on the interaction between
Ce sites, among a magnetically ordered state, a Fermi liquid state, 
a superconducting state, and so on in which the
entropy vanishes at zero temperature.

In these lattice fermion systems, which we call the $f^{2}$ lattice
hereafter, 
the itinerant state is also possible if the hybridization is large
enough.
Thus both the localized f-electron picture and the band picture can be
a starting point to understand the actual compounds with $f^2$
configuration. 
The most interesting situation occurs when the energy scale of the CEF
singlet state is comparable to that of the itinerant state. Then both
states compete for the stability.  

Suppose we have the CEF singlet as the ground state of the $f^2$
lattice, but its energy is only a little lower than the itinerant
state.
If the itinerant state is metallic,  the entropy increases linearly as
temperature increases.  The temperature scale here is the Kondo
temperature,  and is related to the large density of states at the
Fermi surface.
On the other hand increase of the f-electron part of the entropy in
the CEF state follows the exponential law, and is much less
significant in a low temperature range.
Thus there is a possibility for a phase transition to occur 
from the CEF singlet state to the itinerant state
as temperature increases in the $f^{2}$ lattice system. 
Even if the itinerant state is a Kondo insulator,  the entropy can
increase more rapidly than that in the CEF state since the energy gap
decreases with temperature. 
In the latter case the system changes from a metal to an insulator as
temperature increases.
This is opposite to the case of the Mott transition where the
low-temperature phase is an insulator.

The purpose of this paper is to demonstrate the presence of a
phase transition between the CEF 
singlet and itinerant states in the $f^{2}$ lattice system at zero and
finite temperatures. As the first step to explore an $f^{2}$ lattice 
system, we take the simplest possible approach and apply the mean 
field approximation with account of both itinerant and localized 
characters of f electrons. 
The plan of the paper is as follows: In the next section, we introduce
the model and derive the mean-field equations.  The same approximation
scheme is applied to the two-impurity Kondo system with $f^1$
configuration at each site in Section 3.  It turns out 
helpful to compare the electronic states of both models.   
The relative stability of different phases in both the $f^2$ lattice
system and the two-impurity Kondo system is studied in Section 4.
Presence of the phase transition at finite temperatures is
demonstrated.   The final section is devoted to discussion
of results  with attention to possible experimental relevance.

\section{Model and Mean-Field Equations}

We introduce an $f^{2}$ lattice model as follows:
\begin{eqnarray}
H &=& \sum_{{\bf k}\sigma}\sum_{\nu=\alpha,\beta}\varepsilon_{
{\bf k}\nu}c^{\dagger}_{{\bf k}\nu\sigma}c_{{\bf k}\nu\sigma}
\nonumber \\
& &
 + J \sum_{i}\sum_{\nu=\alpha,\beta}\sum_{\mu=1,2}
\left( {\bf S}^{\rm f}_{i\mu}\cdot{\bf S}^{\rm c}_{i\nu}
 + \frac{1}{4}n^{\rm f}_{i\mu}n^{\rm c}_{i\nu} \right)
\nonumber \\
& &
 + I \sum_{i}
\left( {\bf S}^{\rm f}_{i1}\cdot{\bf S}^{\rm f}_{i2}
 + \frac{1}{4}n^{\rm f}_{i1}n^{\rm f}_{i2} \right), 
\label{eq:hamil}
\end{eqnarray}
where $i$ is the site index, and $\nu$ and $\mu$ are channels of
conduction and f electrons, respectively. 
We express the CEF singlet and triplet using a pseudo-spin operator
of f electrons for each channel:
${\bf S}^{\rm f}_{i\mu}=(1/2)\sum_{\gamma\delta}
f^{\dagger}_{i\mu\gamma}
\vec{\sigma}_{\gamma\delta}f_{i\mu\delta}$, where 
$\vec{\sigma}$ is the vector composed of Pauli matrices. 
The spin operator of conduction electrons is given by
$\ {\bf S}^{\rm c}_{i\nu}=
(1/2)\sum_{\gamma\delta}c^{\dagger}_{i\nu\gamma}
\vec{\sigma}_{\gamma\delta}c_{i\nu\delta}$. 
In eq. (\ref{eq:hamil}), the second term with 
$J>0$  gives antiferromagnetic interaction 
between f and conduction electrons on each site. 
In the presence of the potential scattering term 
$(1/4)n^{\rm f}_{i\mu}n^{\rm c}_{i\nu}$, 
the Kondo scale $T_{\rm K}=D\exp(-1/J\rho_{\rm c0})$ is reproduced
correctly in the mean-field approximation. 
Here, $D$ is a half width of a conduction band and $\rho_{\rm c0}$ 
is the  density of states per spin of conduction electrons at the
Fermi 
level. 
We note that the sum of the spin exchange and potential scattering
terms is half of the permutation operator whose eigenvalue is 1 for
the quasi-spin triplet and is $-1$ for the singlet.
The last term with $I>0$ in eq. (\ref{eq:hamil}) 
represents the CEF splitting. This splitting $I$ is also correctly
given by the mean-field approximation due to the term 
$(1/4)n^{\rm f}_{i1}n^{\rm f}_{i2}$. 
The restriction $n^{\rm f}_{i\mu}=1$ is imposed 
on eq. (\ref{eq:hamil}) to simulate the strong Coulomb repulsion
between 
f electrons.

We take a mean field as 
\begin{eqnarray}
V_{i\mu\nu}= -\frac{J}{2}\sum_{\sigma}
\langle f^{\dagger}_{i\mu\sigma}c_{i\nu\sigma} \rangle
\label{eq:f2V}
\end{eqnarray}
for $\mu=1,2$ and $\nu=\alpha,\beta$.  This mean field 
represents the fictitious hybridization between f and conduction
electrons.   
We say "fictitious hybridization" in the sense that the real
hybridization is absent in the model with fixed occupation of f
states.
However we neglect  in this paper the phase fluctuation which makes
the mean field vanish.  The physical motivation for the neglect will
be discussed in the final section.
Another mean-field is given by 
\begin{eqnarray}
R_{i}=\frac{I}{2}\sum_{\sigma}
\langle f^{\dagger}_{i1\sigma}f_{i2\sigma} \rangle
\label{eq:f2R}
\end{eqnarray}
which expresses the mixing between two f orbitals on each site.  
This mixing gives rise to bonding-antibonding splitting of localized
levels. 
In eqs. (\ref{eq:f2V}) and (\ref{eq:f2R}) 
we assume that the mixing is allowed only for the same
spin directions. 
The Lagrange multiplier terms 
\begin{eqnarray}
-\sum_{i}\sum_{\mu=1,2}\lambda_{i\mu}(n^{\rm f}_{i\mu}-1)
\nonumber
\end{eqnarray}
are added to eq. (\ref{eq:hamil}) to enforce the constraints on the
number of f electrons. 
In the mean-field approximation 
the number of f electron per site and channel is fixed only as
average.  
Therefore care is necessary about spurious charge fluctuations
included.
We discuss this aspect of the mean-field theory again in the final
section of the paper.

Assuming equivalence of different sites and channels, we put 
$ \varepsilon_{{\bf k}\nu}=\varepsilon_{\bf k},$ $  
V_{i\mu\nu}=V_{\mu\nu}=|V|\exp(-i\theta_{\mu\nu}),$ 
$R_{i}=R$ 
and $\lambda_{i\mu}=\lambda$. 
Setting the origin of energy at the Fermi level, 
we write the free energy per site as 
\begin{eqnarray}
F &=& -\frac{T}{N_{\rm s}}\sum_{\sigma}
\sum_{{\bf k},\omega_{n}}{\rm tr}\ln
\left[\frac{M({\bf k},i\omega_{n})}{T}\right]  
\nonumber \\
& & +2\left(J+\frac{I}{4}
-\tilde{\varepsilon}_{\rm f}\right)+\frac{8\Delta}{J\pi\rho_{\rm c0}}
+\frac{2|R|^{2}}{I},
\label{eq:Free}
\end{eqnarray}  
where $N_{\rm s}$ is the total number of the sites,
$\tilde{\varepsilon}_{\rm
f}=J+I/4
-\lambda$ and $\Delta=\pi\rho_{\rm c0}|V|^{2}$. The $2 \times 2$
matrix $M$ with Matsubara frequency $i\omega_n$ has components
\begin{eqnarray} 
M_{\mu\mu}&=&-i\omega_{n}+\tilde{\varepsilon}_{\rm f}+
\frac{2|V|^{2}}
{i\omega_{n}-\varepsilon_{\bf k}} \ \ \  (\mu=1,2), 
\\ 
M_{12}&=&-R+\frac{2|V|^{2}}
{i\omega_{n}-\varepsilon_{\bf k}}\Xi, 
\label{eq:xi} \\
M_{21}&=&-R^{\ast}+\frac{2|V|^{2}}
{i\omega_{n}-\varepsilon_{\bf k}}\Xi^{\ast}, 
\end{eqnarray}
where 
$\Xi=\{\exp(-i\phi_{\alpha})+\exp(-i\phi_{\beta})\}/2$ 
with 
$\phi_{\nu}=\theta_{1\nu}-\theta_{2\nu}$ ($\nu=\alpha,\beta$).
Here we represent the magnitude of the
hybridization between different channels of f 
electrons via conduction electrons 
by the parameter $\Xi$.  
Let us take the bases in which the on-site hybridization between f and
conduction electrons occurs only with the same channel. 
Even for this case, the intersite hybridization between the channels 1
and 2 of f electrons can occur.
This is because the point group symmetry around each site is not 
 relevant to intersite interactions.  From a detailed analysis we find
that $R$ and $\Xi$ can be chosen real. 
Then we can assume $-1\le\Xi\le1$ in the following.

We derive the mean-field equations by requiring the free 
energy to be stationary against variation of $\Delta, 
\tilde{\varepsilon}_{\rm f}$ and $R$.
As a result three characteristic states appear: First the Kondo state
is the itinerant state where f-electrons hybridize with  
conduction electrons ($\Delta\ne0$, $R=0$).  Secondly 
the CEF state is the localized state where f-electrons form the
singlet of quasi-spins at each site ($\Delta=0$, $R\ne0$).  Thirdly 
the mixed state has a character interpolating 
between the Kondo and CEF states ($\Delta\ne0$, $R\ne0$). 
For the CEF state  we always have the solution 
$\Delta=\tilde{\varepsilon}_{\rm f}=0$
and 
$R=I\{f(-R)-f(R)\}/2$.
Here, $f(w) =1/\{\exp(w/T)+1\}$ is the Fermi distribution function
with energy $w$.

In the following we consider the case where the number $N_{\rm c}$ of
conduction electrons is twice the number of lattice sites, and the
conduction bands without hybridization have constant density of states
$\rho_{\rm c0}=1/(2D)$ between the band edges $\pm D$.
Then the system has the insulating ground state if f electrons form
energy bands,   
since f and conduction electrons have the hybridization gap at the
Fermi level. 
This is called the Kondo insulator. 
If f electrons are localized, on the other hand, the Fermi level lies
in the middle of the conduction band, and the system becomes metallic.
For the Kondo and mixed states,  each conduction band is split into
two pieces with a band gap between them.  
We use the notation 
$ E_{\zeta}=\tilde{\varepsilon}_{\rm f}-(-1)^{p(\zeta)}R$, and 
$\Xi_{\zeta}=1+(-1)^{p(\zeta)}\Xi$ where 
$p({\rm a})=1$, $p({\rm b})=0$ with $\zeta={\rm a},{\rm b}$. 
The band edges after splitting are given by  
\begin{eqnarray}
D_{1\zeta}&=&\frac{-D+E_{\zeta}}{2}-\frac{D+E_{\zeta}}{2}
\sqrt{1+\frac{8\Delta\Xi_{\zeta}}{\pi\rho_{\rm c0}(D+E_{\zeta})^{2}}},
\nonumber \\
D_{2\zeta}&=&\frac{D+E_{\zeta}}{2}-\frac{D-E_{\zeta}}{2}
\sqrt{1+\frac{8\Delta\Xi_{\zeta}}{\pi\rho_{\rm c0}(D-E_{\zeta})^{2}}},
\nonumber \\
D_{3\zeta}&=&\frac{-D+E_{\zeta}}{2}+\frac{D+E_{\zeta}}{2}
\sqrt{1+\frac{8\Delta\Xi_{\zeta}}{\pi\rho_{\rm c0}(D+E_{\zeta})^{2}}},
\nonumber \\
D_{4\zeta}&=&\frac{D+E_{\zeta}}{2}+\frac{D-E_{\zeta}}{2}
\sqrt{1+\frac{8\Delta\Xi_{\zeta}}{\pi\rho_{\rm c0}(D-E_{\zeta})^{2}}}.
\nonumber 
\end{eqnarray}
Then the mean-field equations are given by 
\begin{eqnarray}
\left(
\int_{D_{\rm 1b}}^{D_{\rm 2b}}+\int_{D_{\rm 3b}}^{D_{\rm 4b}}
\right)
dw \frac{4\Delta\Xi_{\rm b}}{\pi(w-E_{\rm b})^{2}}f(w) &=&
1+\frac{2R}{I}, 
\label{eq:MF1} \\
\left(
\int_{D_{\rm 1a}}^{D_{\rm 2a}}+\int_{D_{\rm 3a}}^{D_{\rm 4a}}
\right)
dw \frac{4\Delta\Xi_{\rm a}}{\pi(w-E_{\rm a})^{2}}f(w) &=&
1-\frac{2R}{I}, 
\label{eq:MF2} \\
\left(
\int_{D_{\rm 1b}}^{D_{\rm 2b}}+\int_{D_{\rm 3b}}^{D_{\rm 4b}}
\right)
dw\frac{\Xi_{\rm b}}{w-E_{\rm b}}f(w)  
\hspace{0.6cm} & & 
\nonumber \\ 
+
\left(
\int_{D_{\rm 1a}}^{D_{\rm 2a}}+\int_{D_{\rm 3a}}^{D_{\rm 4a}}
\right)
dw\frac{\Xi_{\rm a}}{w-E_{\rm a}}f(w) 
&=&-\frac{2}{J\rho_{\rm c0}}.  
\label{eq:MF3}
\end{eqnarray}
We solve eqs.(\ref{eq:MF1})-(\ref{eq:MF3}) 
for various values of dimensionless parameters 
$I/T_{\rm K}$ and $\Xi$ at zero and finite temperatures. 
We note that the mixed state is metallic because the condition
$R\Delta\ne0$ requires a finite density of states between bonding and
antibonding f levels.

\section{Two-Impurity Kondo System}

In the course of understanding the  electronic state in the $f^2$
lattice system,  a necessary step is to clarify the difference from
the $f^2$ impurity system.
This impurity system is further related to the two-impurity Kondo
system with $f^1$ configuration at each site.  
Namely, the $f^2$  impurity is considered as the short-distance limit
of two Kondo impurities.
Fortunately we have detailed knowledge about the two-impurity system
by the mean-field theory \cite{JKM}, the Quantum Monte Carlo
 \cite{Hirsch}, and  the numerical renormalization
group \cite{Jones,Sakai}.
In this paper we derive the ground state and the free energy in the
same level of approximation as is used for the $f^2$ lattice system.
Then by comparing the electronic state of the $f^2$ lattice system and
that of the impurity system, we obtain information about the influence
of the lattice periodicity. 

In ref. \cite{JKM} the two-impurity Anderson model was solved by the
mean-field theory. 
It was shown that the intersite hybridization gives smooth change from
the limit of two independent Kondo states (Kondo pair) to the pair
singlet state as the intersite interaction increases.  
Physically we expect the same situation even though the occupation of
f electrons at each site is very close to unity.  
The limiting case is described by the two-impurity Kondo model.
Although the Anderson lattice model is more general than the Kondo
lattice model, the CEF state is harder to treat in the mean-field
theory.  Since we have adopted the Kondo lattice model with $f^2$
configurations,  we need to solve the two-impurity Kondo model for
comparison.  

The two-impurity model is given by
\begin{eqnarray}
H_{\rm 2imp}&=&\sum_{{\bf k}\sigma}\varepsilon_{\bf k}c^{\dagger}
_{{\bf k}\sigma}c_{{\bf k}\sigma}
+J\sum_{j=1}^{2}\left({\bf S}^{\rm f}_{j}\cdot{\bf S}^{\rm c}_{j}
+\frac{1}{4}n^{\rm f}_{j}n^{\rm c}_{j}\right)
\nonumber \\
&+&I\left({\bf S}^{\rm f}_{1}\cdot{\bf S}^{\rm f}_{2}
+\frac{1}{4}n^{\rm f}_{1}n^{\rm f}_{2}\right), 
\label{eq:2impH}
\end{eqnarray}
where $J, I>0$ and $j (=1,2)$ labels sites of f electrons.
There is only a single conduction band since even in this case
different screening channels are present around each impurity.  
We take the mean-fields in the form analogous to the 
$f^{2}$ lattice system:  One is given by 
\begin{eqnarray}
V_{j}=-\frac{J}{2}\sum_{{\bf k}\sigma}e^{i{\bf k}\cdot{\bf r}_{j}}
\langle f^{\dagger}_{j\sigma}c_{{\bf k}\sigma} \rangle
\nonumber
\end{eqnarray}  
which 
represents the fictitious hybridization between f and conduction
electrons at each site. 
The other is given by 
\begin{eqnarray}
R=\frac{I}{2}\sum_{\sigma}
\langle f^{\dagger}_{1\sigma}f_{2\sigma} \rangle
\nonumber
\end{eqnarray} 
which expresses the mixing between two f electrons.  
As before the Lagrange multiplier terms 
\begin{eqnarray}
-\sum_{j=1,2}\lambda_{j}(n^{\rm f}_{j}-1)
\nonumber
\end{eqnarray}
are added to eq. (\ref{eq:2impH}) to enforce the constraints on the
number of f electrons. 

There are two dimensionless parameters $A, B$ which 
represent the intersite hybridization effect via conduction electrons.
Namely we define  
\begin{eqnarray} 
|V|^{2}\sum_{\bf k}
\frac{e^{i{\bf k}\cdot{\bf r}}}
{i\omega_{n}-\varepsilon_{\bf k}}
= [B+iA{\rm sgn}(\omega_{n})]\Delta.
\label{eq:AB}
\end{eqnarray}
Here, the magnitudes of $A$ and $B$  depend on 
both the distance ${\bf r}$ between f sites and the band structure
of 
conduction electrons, but they are always less than unity \cite{JKM}. 
The parameter $A$ causes asymmetry in the density of states of bonding
and antibonding f states: 
In the case $R>0$,  with $A>0$ $(<0)$ the density of 
states of the bonding states becomes wider (narrower) than that of the
antibonding states. 
On the other hand the parameter $B$ controls the splitting between
bonding and antibonding f states. 
The left-hand side of eq. (\ref{eq:AB}) is analogous to the 
term with $\Xi$ in eq. (\ref{eq:xi}) if one interchanges sites 
in the former with channels in the latter. 
In the two-impurity system there are three characteristic states: 
The
first is the Kondo-pair state where the Kondo effect occurs
independently at 
each site ($\Delta\ne0, R=0$);
the second is the pair-singlet state 
where the pair-singlet of f electrons is formed without help of 
conduction electrons ($\Delta=0, R\ne0$); 
the third is the mixed state which interpolates the above two states
($\Delta\ne0,  R\ne0$).

\section{Stability of Itinerant and Localized States}
\subsection{Zero temperature}

We have solved the mean-field equations numerically at zero
temperature both for the $f^2$ lattice and the two-impurity systems.
Table \ref{table:para} summarizes the parameters used in the
calculation. 
Figure \ref{fig:T0phase}(a) shows the ground-state energy
per site in the $f^{2}$ lattice system at zero temperature. 
The origin of energy is taken to be that of the Fermi sea without
f electrons.
The abscissa represents the bare CEF splitting in units of  
$T_{\rm K}$. 
The notations $E_{\rm K}$, $E_{\rm CEF}$ and $E_{\rm mix}$ 
represent the ground-state energies of Kondo, 
CEF and mixed states, respectively. 
The effect of intersite hybridization $\Xi$ depends only on its
absolute value.  Thus results of $E_{\rm mix}$  with $\Xi=0$ and
$\Xi=0.4$ are shown as representative cases in Fig.
\ref{fig:T0phase}(a).   

We find that $E_{\rm mix}$ 
is larger than $E_{\rm K}$ and $E_{\rm CEF}$ for all  
combinations of parameters $I/T_{\rm K}$ and $\Xi$.  
Therefore the change from the Kondo state to the CEF state
occurs discontinuously at the critical point $I/T_{\rm K}=4$. 
The mixed state which would have interpolated the Kondo and CEF states
smoothly is not stabilized actually;  
with increasing intersite hybridization,  the mixed state with energy
$E_{\rm mix}$ becomes larger in the mean-field theory.  
This is seen by the fact that $E_{\rm mix}$ with $\Xi=0.4$ is larger
than that with $\Xi=0$ in Fig. \ref{fig:T0phase}(a). 
The reason is the following:
 If $\Xi>0$ and $R>0$, the density of states of 
the bonding f states has a larger width than that of the antibonding f
states. 
The ground-state energy is given by the sum of  
single-particle energies of occupied states.  Namely, we integrate the
total density of states multiplied by $w$ from $-\infty$ to 0.
Since the integral without $w$ is fixed by the number conservation,
the total energy increases by the asymmetry induced by $\Xi$.
Similarly in the case of $\Xi<0$ and  $R>0$, the asymmetry of the
density of states in the opposite direction increases the energy
again. 

For comparison, Fig.\ref{fig:T0phase}(b) shows the
ground-state energy of f electrons in the two-impurity system 
at zero temperature. 
We have tried various values of $A$ and confirmed that $A$ does not
influence the relative stability of the phases.  On the contrary the
value of $B$ drastically affects the ground state. 
Hence, we fix $A=-0.2$ and vary $B$ as a free parameter.
We note that if $B=0$, $E_{\rm mix}$ is larger than both $E_{\rm K}$
and $E_{\rm ffpair}$ for any value of $I/T_{\rm K}$.  
This situation is analogous to that in the $f^2$ lattice system.
As a result an abrupt change from the Kondo-pair state to the
pair-singlet state occurs 
as $I/T_{\rm K}$ is increased.  
At the critical point of $I/T_{\rm K}=2.5$, the two kinds of singlet
states are degenerate.
Thus one observes the divergence of physical 
quantities such as the susceptibility and the specific heat
coefficient. 
We have checked that this level-crossing behavior remains the same as
long as $|B|<1/\pi$.
On the contrary, if the hybridization effect is large ($1/\pi<|B|<1$),
the Kondo-pair state connects continuously with the pair-singlet state
through the mixed state.
In this case no divergence occurs.    
This is shown in Fig.\ref{fig:T0phase}(b) by the result that $E_{\rm
mix} (B=0.4)$ is lower than both $E_{\rm K}$ and $E_{\rm ffpair}$ for
any value of $I/T_{\rm K}$. 
These results obtained in the mean-field approximation 
agree with those in refs. \cite{Hirsch,Sakai,JKM} for the two-impurity
Anderson model.   

We note that $A$ controls asymmetry 
of the density of states of f electrons with respect to the Fermi
level, just as $\Xi$ does in the $f^{2}$ lattice system. 
We have confirmed that $E_{\rm mix}$ becomes larger as $|A|$ increases
with $B$ being fixed. 
Thus the parameter $A$ plays the same role as $\Xi$ in the $f^{2}$
lattice system.
However  in the lattice system we do not have the parameter
corresponding to $B$ in the impurity system. 
Hence there is no stable mixed state in the $f^{2}$ lattice  system in
our calculation.  
Mathematically the ineffectiveness $\Xi$ in stabilizing the mixed
state comes from the absence of $\bf k$-summation in the 
self-energy of the $f^{2}$ lattice system.

\subsection{Finite Temperature}

The mean-field equations are solved numerically also at finite
temperatures, and the free energies are derived.
Figure \ref{fig:free}(a) shows temperature dependence of 
free energies per site
for three different states in the $f^{2}$ lattice system: 
Kondo, CEF and mixed states.  
Even though the CEF singlet is the ground state, there is a case
where the itinerant state is realized at higher temperatures. 
We find that the free energy of the mixed state is larger than those
of the other two states for all values of parameters $I/T_{\rm K}$ and
$\Xi$. 
Therefore, the transition between the Kondo and CEF 
phases occurs as a first-order one. 

Figure \ref{fig:free}(b) shows free energies in the two-impurity
system.  The notations 
$F_{\rm K}$, $F_{\rm ff pair}$ and $F_{\rm mix}$  
represent the free energies of the Kondo-pair, 
pair-singlet and mixed states, respectively. 
It is seen that with $B=0$,  $F_{\rm mix}$ is larger than 
$F_{\rm K}$ and $F_{\rm ffpair}$. 
As in the case of zero temperature, the mixed state is not stabilized
as long as $|B|<0.08$ for all values of $I/T_{\rm K}$. 
On the other hand, $F_{\rm mix}$ with $B=0.4$ in Fig.
\ref{fig:free}(b)
is lower than both $F_{\rm K}$ and $F_{\rm CEF}$. 
We have checked that with $0.08<|B|<1$ the mixed state is stabilized
at all temperatures.  

From these results we infer  that 
the parameter $B$ in the two-impurity system plays a decisive role 
also at finite temperatures.
Furthermore in the $f^{2}$ lattice system there is no temperature
region where the mixed state is stabilized.
In other words, the hybridization effect which mediates between Kondo
and CEF phases is ineffective, and the transition occurs 
discontinuously.

\section{Discussions}

\subsection{Comparison with two-impurity systems}

In considering the relevance of the mean-field theory, we first take
the case of impurity systems. 
The physical difference between the two-impurity Kondo and Anderson
models is whether there is
charge fluctuations of f electrons or not.
 Reliable knowledge is available for both models 
from several numerical calculations.  
Computation using the numerical renormalization group
derived a level crossing between the Kondo-pair state and the
pair-singlet state \cite{Jones}.  As a result divergence of the
staggered susceptibility occurs at zero temperature.
On the contrary,  a quantum Monte Carlo calculation for the
two-impurity Anderson model \cite{Hirsch} found continuous behavior in
physical quantities.  This apparent conflict was resolved by Sakai et
al. \cite{Sakai} who identified the origin of the continuous crossover
as the bonding-antibonding splitting of f orbitals.  
In the Kondo model, the splitting is absent because there is no charge
degrees of freedom for f electrons.
Thus the divergent behavior is purely a formal consequence of the
model since there should always be some amount of charge fluctuations
in real systems.

For our purpose of studying a new type of phase transition,  we regard
our model given by eq.(\ref{eq:hamil}) only as a simplified form of
Anderson-type models which are more difficult to analyze by the
mean-field theory.
Then the finite order parameters for various phases are rather to be
regarded as properties of a corresponding Anderson-type model. 
In formally exact treatment of eq.(\ref{eq:hamil}),  all of our order
parameters would vanish identically in contrast with the results of
the mean-field theory.
However by the same reason as explained above for the two-impurity
models, we would rather accept the results of the mean-field theory as
a physically possible consequence for more realistic models.
 
\subsection{Comparison with the $f^1$ lattice system}

It is instructive to take the limits $I\rightarrow 0$ and
$\Xi\rightarrow 0$ in eq.(\ref{eq:hamil}).  Then the system becomes
equivalent to two independent Kondo lattices.  
In the half-filled case the ground state is either the Kondo insulator
or a magnetically ordered phase.  
The latter state is due to the RKKY interaction which is not taken
into account by the mean-field theory.   
We note that the ordered state can be either metallic or insulating
depending on the magnetic structure.  If it is ferromagnetic, the
half-filled conduction band leads to the metallic state.
As one increases $J$ from zero, the ground state should change from
the magnetically ordered state to the insulating one.  Subsequently
the nature of f electrons changes from the localized character to the
itinerant one.
Since the entropy is different in the two phases,  a phase transition
can occur from one phase to the other as a function of temperature.

Let us compare this phase transition with another one which is known
as an artifact of the mean-field theory.  Namely, as temperature
increases in the mean-field theory, the order parameter $\Delta$ of
the Kondo insulating phase decreases continuously to zero around the
temperature $T\sim T_{\rm K}$.  
In the exact theory the local gauge fluctuation washes away the
transition completely.
We emphasize that the possible transition between the Kondo insulator
and the magnetically ordered phase should survive the fluctuation
effect.

Now we consider the case of finite $I$ and $\Xi$.   In the mean-field
theory, the Kondo insulating phase does not feel the effect of $I$ and
$\Xi$.  The resultant state is the same as the direct product of the
two $f^1$ lattices.  Hence the second-order transition around $T\sim
T_{\rm K}$ is again fictitious.  
One can ask at zero temperature how the magnetic state changes as $I$
increases continuously  from 0.   
For small $I$, we have the ordered induced moment which arises by
mixing of the singlet and triplet levels.   
At certain critical value of $I$, the CEF singlet will become more
stable than the induced moment.  
The situation is analogous to the spin chain problem where exchange
interactions of alternating strength form dimers with an excitation
gap.
We note that the f electrons are always localized for any value of
$I$.
Thus the change to the Kondo insulator can occur as a phase
transition, although both states are spin singlets.

\subsection{Effects of charge fluctuation on the phase transition}

In actual $f^{2}$ lattice systems there should always be charge 
fluctuation as discussed above.
Then any exact eigenstate has some amount of hybridization between f
and conduction electrons.  
However,  there can still be two different kinds of hybridized states:
The first one can be reached by perturbation theory with respect to
hybridization.
This state is connected with the localized $f^{2}$ state. 
The other state is the itinerant state which is not accessible by such
perturbation theory.   The latter state instead is simply described by
the band picture of f electrons.
Thus possibility of the phase transition remains even though effects
of charge fluctuations are included.

We note that the transition to the CEF singlet phase is of first
order.  
A first-order transition should be less sensitive to fluctuation
effects than a second-order one.   We plan to check the robustness of
the phase transition by using theories \cite{XNCA,SK,Georges} more
reliable than the mean-field theory.

\subsection{Possible experimental relevance}

With respect to experimental relevance,  we have to consider also the
case where the number of conduction electrons deviates from $2N_{\rm
s}$.
In the Anderson lattice model with dominant $f^2$ configurations, the
itinerant state then has a finite density of states of f electrons 
at the Fermi level, and hence is metallic.
In this case the Fermi level is shifted from the center of the
hybridization gap, and the average occupation of f electrons per site
also deviates from 2.
Thus in reality the transition from the itinerant state to the
localized one is not always an insulator-metal transition. 

Concerning possible relevance of our theory, we mention two uranium
compounds: UNiSn and 
$\rm URu_{2}Si_{2}$.  In the former case, the insulating state at high
temperatures changes via a first-order transition into a metallic
state at $T= 43$ K.  In contrast to our model, however, the metallic
state shows the antiferromagnetic order \cite{Aoki}.
As long as the localized picture applies to the low-temperature phase,
the driving force of the transition may be similar to the one
discussed in this paper.  
It is necessary to include the induced moment for more detailed
analysis of UNiSn. 
In the latter case of $\rm URu_{2}Si_{2}$,  the CEF singlet model
accounts for gross features of highly anisotropic susceptibility and 
metamagnetic transition \cite{Niewen}.  
The high-temperature phase is metallic showing the Kondo effect in the
resistivity.  
A clear anomaly in the specific heat is observed at temperature
$T_{\rm 0}=17.5$ ${\rm K}$ \cite{schlabitz}, and 
the resistivity shows the metallic behavior also below $T_0$.
By neutron scattering \cite{brohorm} the antiferromagnetically ordered
magnetic moments were observed below $T_{\rm 0}$.
The magnitude of the moment is only $ 0.04\mu_{\rm B}$ which is
smaller by two orders of magnitude than the usual magnitude observed
in similar compounds ${\rm
UT_{2}Si_{2} (T=Pd,Rh)}$. 
Moreover, the growth of moments with decreasing temperature does not
follow the mean-field behavior.  
Strangely, the NMR does not probe the internal field below $T_{\rm 0}$
 \cite{Kohori}.
Thus there is a possibility that the apparent antiferromagnetism is
not a true long-range order, but due to very slow fluctuation of U
moments.

In any case the specific heat jump at $T_{\rm 0}$  is too large to be
accounted for by the tiny magnetic moment.   
Thus, the proper order parameter in this ordered phase 
remains to be identified \cite{Gorkov,SA,Walker,Mason}. 
We note that the inelastic neutron scattering \cite{brohorm} probed a
feature which looks like a CEF excitation below $T_{\rm 0}$. 
This fact may be a key to identify the order parameter. 
The phase transition seems to be of second order.

In summary, we have shown that the phase transition from the itinerant
state to the CEF singlet state occurs as temperature decreases in the
$f^{2}$ lattice system in the frame of the mean-field theory. 
Properties of the $f^{2}$ lattice system were discussed 
in comparison with the two-impurity system at zero and finite
temperatures. 
We suggest that the competition between localized and itinerant
states of f electrons is the fundamental driving force for phase 
transitions in some uranium compounds such as UNiSn and $\rm
URu_{2}Si_{2}$.
It remains to see to what extent the fluctuation effect beyond the
mean-field theory affects the phase transition.

\newpage 
\begin{figure}[h]
\caption{The ground-state energies of (a) the $f^{2}$ lattice system,
and
(b) the two-impurity system.  
In (a), $E_{\rm K}, E_{\rm CEF}$ and $E_{\rm mix}$ correspond to the
Kondo (itinerant) , CEF (localized) and mixed states, respectively,
and $\Xi$ is a parameter to characterize the strength of intersite
hybridization.  
In (b), $E_{\rm K}, E_{\rm ff pair}$ and $E_{\rm mix}$ show the
energies of 
Kondo-pair, pair-singlet and mixed states, respectively.
The parameter $B$ ($|B|\le1$) characterizes the strength of intersite
hybridization effect.
Another hybridization parameter $A$ is fixed to be  $-0.2$ (see
text).}  
\label{fig:T0phase}
\end{figure}

\begin{figure}[h]
\caption{Free energies of (a) the $f^{2}$ lattice system and 
(b) the two-impurity system.   In (a), $F_{\rm K},F_{\rm CEF}$ and
$F_{\rm mix}$ indicate the Kondo (itinerant), CEF and mixed states,
respectively.
In (b), $F_{\rm K}, F_{\rm ff pair}$ and $F_{\rm mix}$ indicate the 
Kondo-pair, pair-singlet and mixed states, respectively.
The parameter $A$ is fixed to be $-0.2$ as in Fig.1.}
\label{fig:free}
\end{figure}

\newpage
\begin{table}[h]
\caption{Parameters for calculation.}
\begin{center}
\begin{tabular}{c c c c} 
\hline
$D$ & $T_{\rm K}$ & $I$ & $\Xi$\\ 
\hline
$10^{4}$ [K] & $1$ [K] & $0\sim10$ [K] & -1$\sim$1 \\
\hline 
\end{tabular}
\end{center}
\label{table:para}
\end{table}

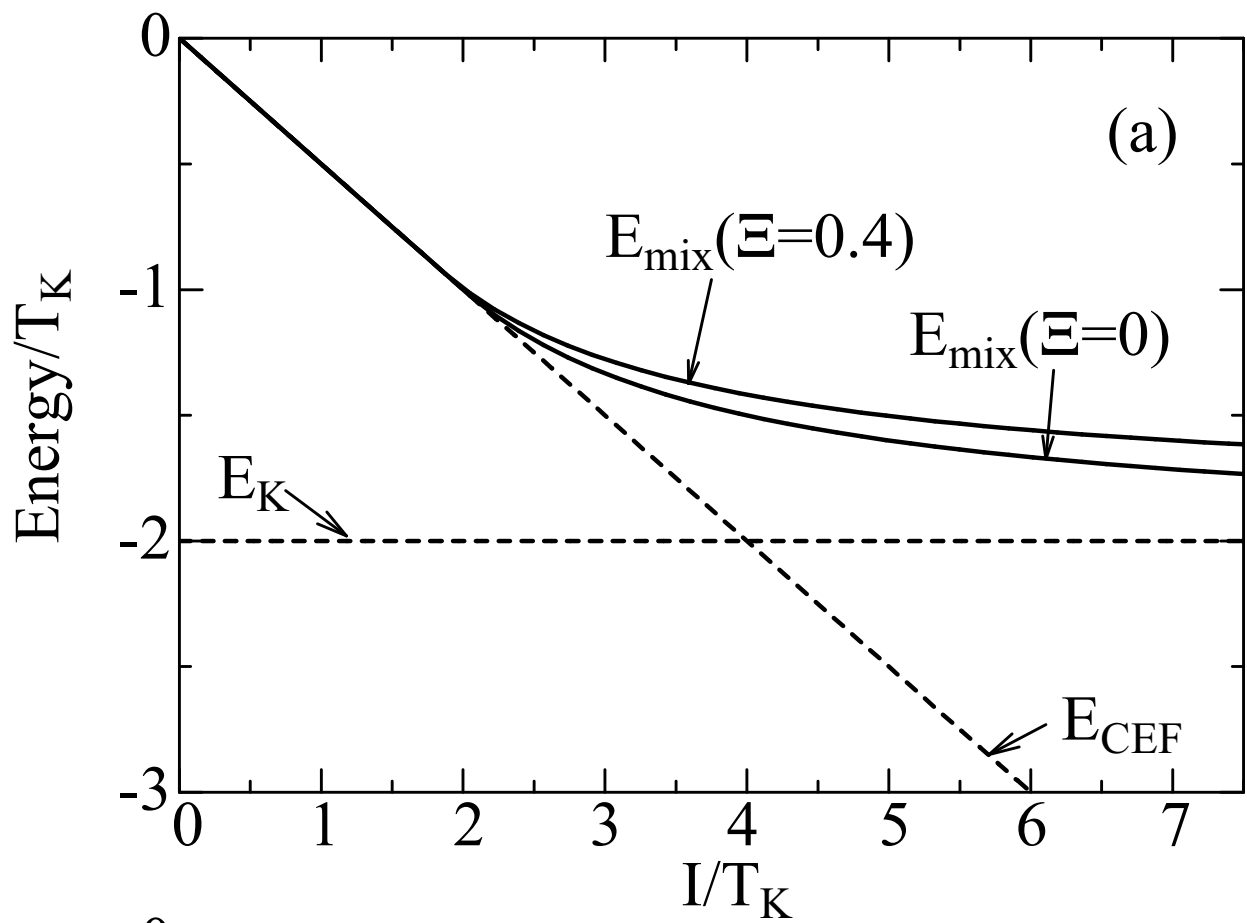
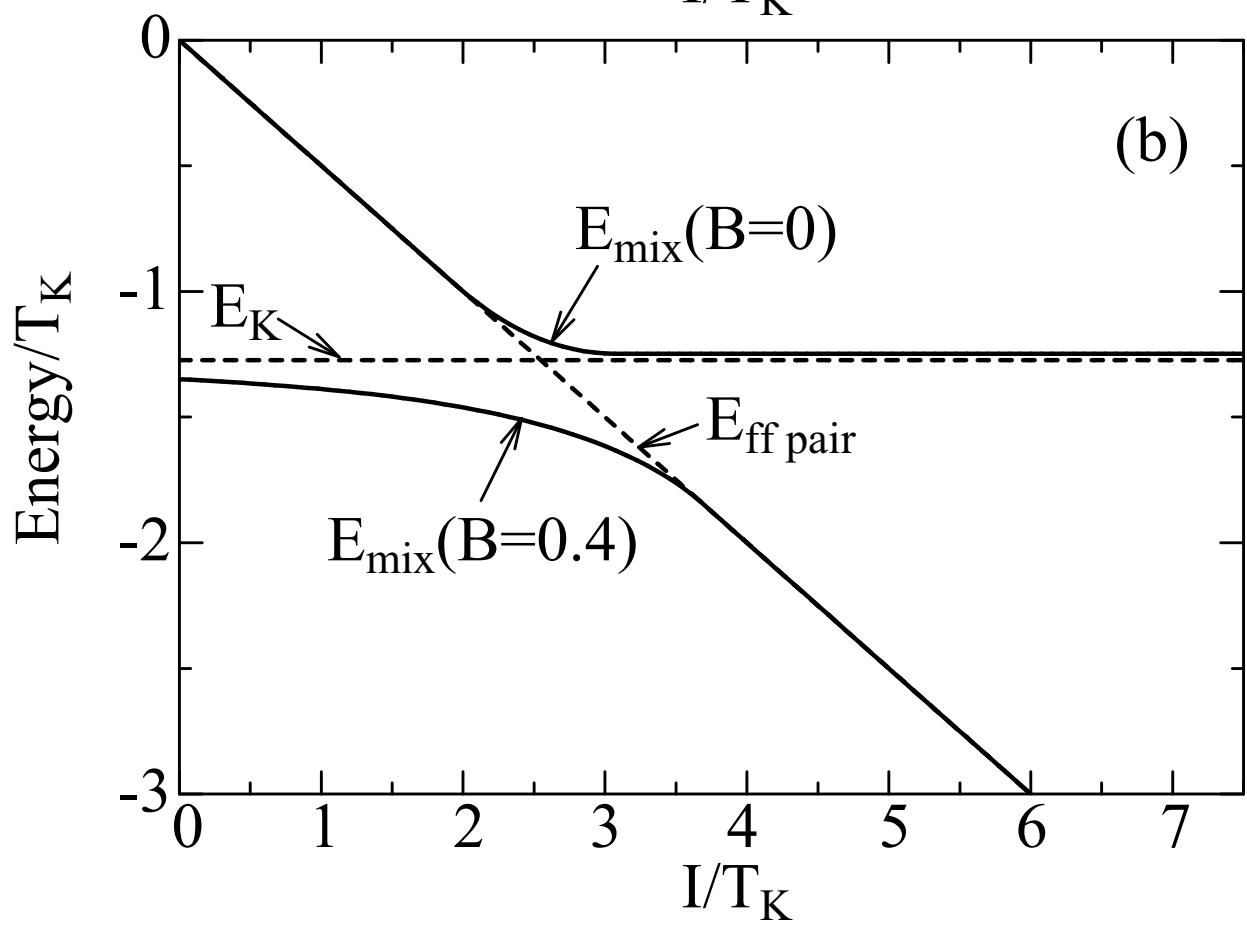
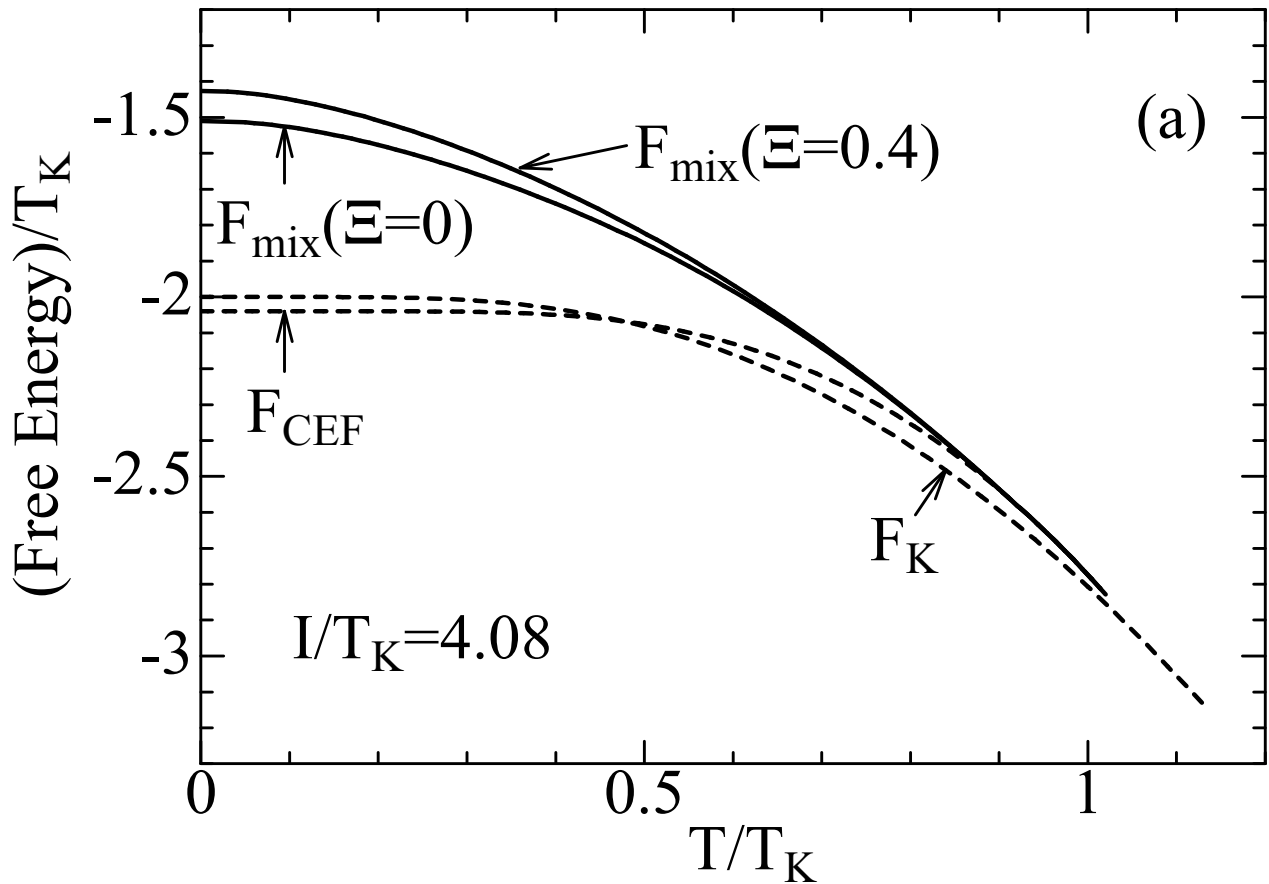
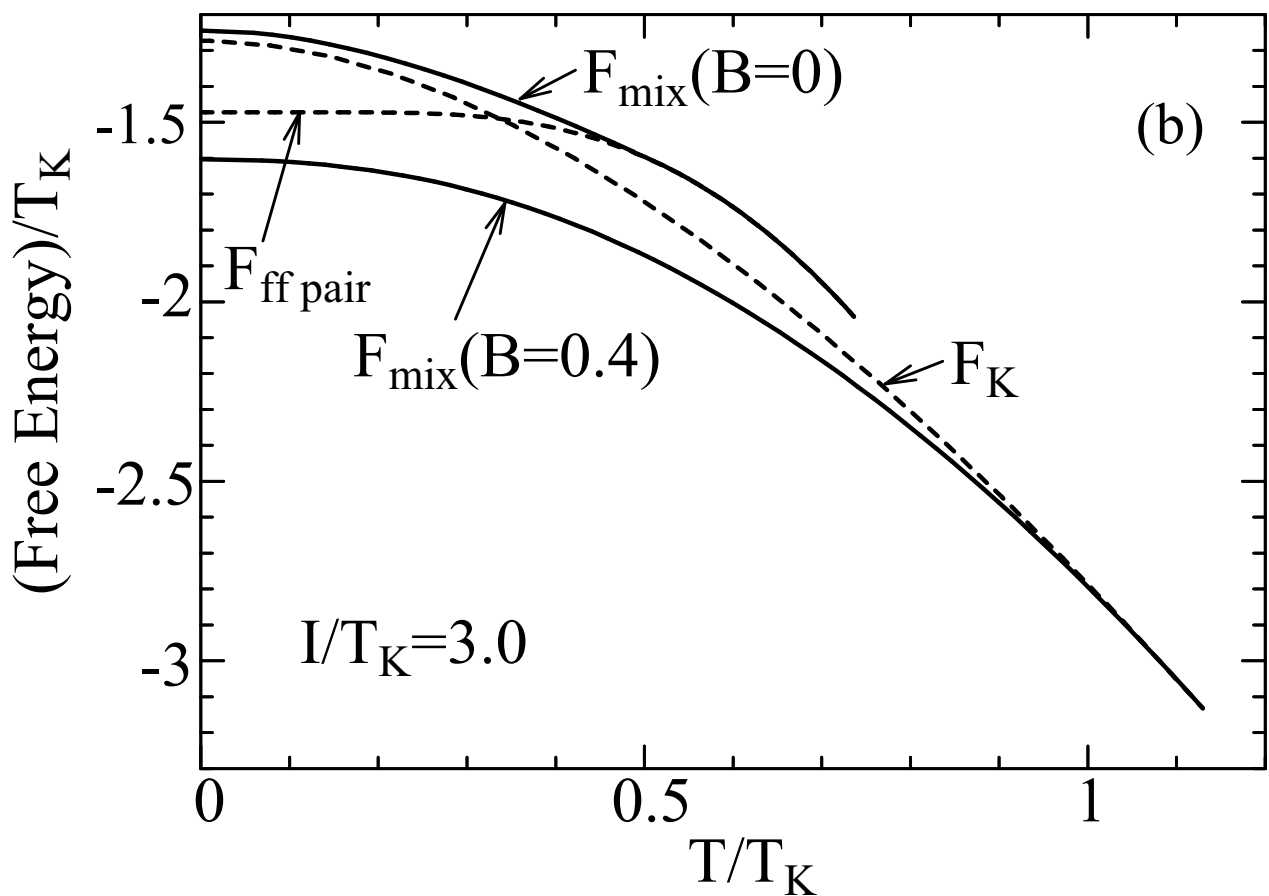
\end{document}